\documentclass{ifacconf}

\usepackage{graphicx}      
\usepackage{natbib}
\usepackage{amsfonts}
\usepackage{graphicx}
\usepackage{subfigure}
\usepackage{epstopdf}
\usepackage{amssymb}

\newtheorem{proposition}{\textbf{Proposition}}
\begin{document}
\begin{frontmatter}

\title{Event triggering control for dynamical systems with designable minimum inter-event time}

\thanks[footnoteinfo]{This work was supported by the National Science Foundation of China under
Grants 61703130, by the Scientific Research Fund of Education Department of Yunnan Province under
Grants 2019J0009.}

\author[First]{Xing Chu}
\author[Second]{Na Huang}
\author[Third]{Zhiyong Sun}

\address[First]{National Pilot School of Software, Yunnan University, Kunming, Yunnan, P.R.China (e-mail: chx@ynu.edu.cn).}
\address[Second]{School of Automation, Hangzhou Dianzi University, China \\(e-mail: huangna@hdu.edu.cn).}
\address[Third]{Department of Automatic Control, Lund University, Lund, Sweden \\(e-mail: sun.zhiyong.cn@gmail.com)}

\begin{abstract}                
This paper presents a class of event-triggering rules for dynamical control systems with guaranteed positive minimum inter-event time (MIET). We first propose an event-based function design with guaranteed control performance under a clock-like variable for general nonlinear systems, and later specify them to general linear systems. Compared to the existing static and dynamic triggering mechanisms, the proposed triggering rules hold the robust global event-separation property, and can be easily implemented on practical digital platform. Namely, it is shown that the minimum inter-event time can be flexibly adapted to the various hardware limitations. Finally, several numerical  simulations are given to illustrate the theoretical results.
\end{abstract}

\begin{keyword}
Event-triggered control, minimal inter-event time, dynamic triggering mechanism, robustness analysis.
\end{keyword}

\end{frontmatter}

\section{Introduction}
The emerging application of wireless communication technique in the conventional feedback control systems creates a new type of  control systems that is often called networked control systems \cite{ZHY15}. It offers a variety of benefits including flexibility, maintainability and cost reduction of automation process. Nevertheless, it also produces considerable design challenges like the extra energy consumption, and additional constraints  in the closed-loop systems, such as communication bandwidth and control update frequency. In recent years, it is shown that event-triggered control paradigm is a promising solution compared to the commonly used time-triggered ones to update control when needed. Instead of continuous sampling and communication, event-triggered control scheme can determine the time when the state information is needed to sample and send to the control law based on certain triggering rules; see \cite{AB02, T07}. Often in event-triggered control, a key issue is the exclusion of possible Zeno triggering behavior. In this paper, we will present a practical event triggering mechanism that guarantees positive minimum inter-event time (MIET), thus excluding the Zeno triggering for an event-triggered control system.

Event-triggered control has received considerable attention in recent years, and here we review some key developments in this field.
The pioneer paper  \cite{T07} initialized a general event-based scheduling control for general nonlinear systems, and a typical design and analysis framework for  event-triggered control systems was formally proposed.   Generally speaking, the static triggering rule in \cite{T07} and many subsequent papers often consists of system state variables. An event-triggered control scheme is proposed and analyzed for perturbed linear systems in \cite{HSV08}, where the sampling is performed only when the tracking/stabilization error is large enough. Another paper \cite{A08} presented the architecture of a general structure for event-triggered control and discussed the relations to nonlinear systems. Interestingly, instead of using zero-order holder in most of event triggering mechanisms, the control input is designed to mimic a continuous feedback between two consecutive event times in \cite{LL10}. Later, another work in \cite{WL11} gave a scheme to postpone the triggering time over previously proposed methods so as to enlarge the MIET. Meanwhile, the model-based control technique and event-triggered mechanism were unified into a single framework to stabilize the uncertain general linear systems in \cite{GA12}. The network induced delays and quantization errors were considered and related results were derived. Also, the synthesis of event-triggering rule and controller for delayed linear systems was investigated and the optimization problem was considered with respect to two kinds of performance indices in \cite{WRG12}, respectively. Typically, in \cite{HDT12} the periodic event-triggered control approach was formulated for general linear networked control systems, in which the MIET is guaranteed to be at least the fixed sampling period. A concept of robust event-separation property was proposed in \cite{BH14}, which shows that some popular event triggering mechanisms do not ensure the event separation property no matter how small the disturbance is. This is an important aspect in the design of triggering function for event-based control system under perturbations. For more development of event-based control and triggering function design, the reader is referred to the   surveys \cite{HJT12, zhang2016overview}. Unfortunately, for all of above work, it is worth mentioning  that the upper bounds of those constructed triggering signals are hard to adjust. Specifically, although the Zeno behavior can be excluded by guaranteeing a strictly positive MIET, the value of MIET cannot flexibly adapt physical limitations of hardwares, which means that those event triggering mechanisms can not be realized on digital platform.

Recently, to improve the results in \cite{T07}, the dynamic triggering mechanism was formally proposed in \cite{G14}. In this framework,  an internal dynamic variable is introduced into the   static triggering rule, which also helps to enlarge the next triggering time instant. However, similar to the static ones, the resulting MIET still can not adapt to the hardware limitations, also the triggering signal is dependent of the real systematical states with specific constraints . For other recent remarkable work of dynamic triggering mechanism, we notice that in \cite{ACM2016} the authors considered the output feedback stabilization problem for the general linear systems with event-triggered sampling and dynamic quantization and that in \cite{BAA18} a new type of event condition was proposed to be dependent on the state difference between the actual system and the nominal undisturbed system, which is triggered when the nominal state is equal to the state of real system. For other interesting results on dynamic event-triggered control, we further refer the readers to the overviews \cite{NGC19,DHG17}.

Based on the above discussion, we notice that under  the static or dynamic triggering mechanisms, the variable range and the evolution rate of the constructed triggering signal cannot be freely designed. In the meantime, the investigation on the robustness issue of dynamic event triggering mechanism is lacking. These motivate the construction of an event triggering mechanism with a designable MIET and a robust global event-separation property. We remark that only in \cite{JC19} that the designable MIET control is discussed, in the context of multi-agent consensus control with single-integrator dynamics.  In this paper, we aim to propose an event-triggered control scheme to realize adjustable positive MIET and guaranteed system convergence. The main contributions of this paper are summarized as below.
\begin{enumerate}
  \item We show new design and analysis approaches for dynamic event triggering mechanism, which are applied to general nonlinear system and also specified to general linear systems.
  \item A freely designable MIET is derived. Compared to the static and dynamic event triggering mechanisms, we can adjust the variable range of the constructed triggering signal regardless of physical limitations of real systematical states.
  \item The proposed dynamic event-triggered strategy ensures the robust global event-separation property under state perturbations.
\end{enumerate}

The rest of this paper is organized as follows. In Section 2, we review two event-triggered schemes in the literature and some preliminary knowledge of event-triggered control, and present the problem formulation.    The design and analysis framework of MIET-designable event triggering mechanism  for nonlinear systems is presented in Section 3. In Section 4, the framework presented in Section 3 is specified to general linear control system, while the robustness of the proposed algorithm is analyzed. In Section 5, two simulation examples are provided to illustrate the effectiveness of the present theoretical results. Finally, remarking conclusions are given in Section 6.\\

{\bf Notations.}
Throughout this paper, $\mathbb{R}$ and $\mathbb{R}^n$ denote the set of real numbers and the $n$-dimensional Euclidean space, respectively. $\mathbb{R}^{+}_{0}$ is the set of non-negative real numbers. The notation $|\cdot|$ refers to the Euclidean norm for vectors and  the induced 2-norm for matrices. The superscript $\top$ denotes vector or matrix transposition. A function $\alpha(r):\mathbb{R}^{+}_{0}\rightarrow\mathbb{R}^{+}_{0}$ is said to be of class $K_{\infty}$ if it is continuous, strictly increasing, $\alpha(0)=0$, and $\alpha(r)\rightarrow+\infty$ as $r\rightarrow+\infty$.

\section{Event-Triggering Mechanism}
We consider the control system of the form
\begin{eqnarray}
\label{nonlinear system}
\dot x=f(x,u), x\in \mathbb{R}^n, u\in \mathbb{R}^m
\end{eqnarray}
with a state feedback law $u=k(x)$ that stabilizes the system. The resulting closed-loop control system is shown below
\begin{eqnarray}
\label{continuous closed-loop system}
\dot x=f(x,k(x))
\end{eqnarray}
Under the state feedback law $u=k(x)$  the state information of the plant should be available and accessed at a continue manner so as to update the input continuously. In event-triggering control, the state information is accessed when necessary and the control input is updated when certain events occur. This results in a discrete-time updated control law $u=-k(x(t_i)), t\in[t_i,t_{i+1})$, where $t_i$ denotes the $i$-th triggering time instant. We note that if $t_{i+1}-t_{i}\rightarrow0$ for some finite time $t_i$, the sampling process becomes impractical (which is termed a Zeno triggering). Therefore, how to guarantee a positive MIET is one of the key issues in the design of event-triggering control.

As in \cite{T07}, we define the   measurement error as $e(t)=x(t_i)-x(t), t\in[t_i,t_{i+1})$. We assume that the  closed-loop control system
\begin{eqnarray}
\label{discrete closed-loop system}
\dot x=f(x,k(x+e)).
\end{eqnarray}
is input-to-state stable (ISS)  with respect to $e(t)$.
There thus exists an ISS-Lyapunov function $V$ for class $K_{\infty}$ functions $\bar\alpha, \underline{\alpha}, \alpha$, and $\gamma$ satisfying the following inequalities
\begin{eqnarray}
\label{ISS}
\underline{\alpha}\leq V(x)\leq\bar{\alpha},\nonumber\\
\dot V(x)\leq-\alpha(|x|)+\gamma(|e|).
\end{eqnarray}
\subsection{Static Event Triggering Mechanism}
The seminal paper  \cite{T07} proposed an event triggering strategy for the system (\ref{discrete closed-loop system}), with the following static triggering rule
\begin{eqnarray}
\label{static triggering rule}
t_{i+1}=inf\{t>t_i|\gamma(|e|)\geq \sigma\alpha(|x|)\}
\end{eqnarray}
In event-triggered control systems,  the interval time of  event triggering  should be designed to satisfy
\begin{eqnarray}
\label{MIET}
t_{i+1}-t_{i}\geq\tau >0, \forall i,
\end{eqnarray}
where $\tau$ is a positive constant on MIET.
It is shown in \cite{T07} that for all initial state $x(0)\to S$ where $S\subset\mathbb{R}^n$ is a compact set containing the origin, there exists $\tau>0$ such that the sequence $t_i$ determined by (\ref{static triggering rule}) satisfies (\ref{MIET}) if $f(\cdot)$, $k(\cdot)$, $\alpha^{-1}(\cdot)$ and $\gamma(\cdot)$ are Lipschitz continuous on compacts and $0<\sigma<1$.


\subsection{Dynamic Event Triggering Mechanism}
This dynamic triggering mechanism \cite{G14} is an extension of the static triggering scheme   through enlarging the variable range of the constructed triggering signal $\gamma(|e|)/\alpha(|x|)$. To explain this point, let us recall the triggering rule (\ref{static triggering rule}) and add a positive item in the right hand side of the inequality
\begin{eqnarray}
\label{}
\frac{\gamma(|e|)}{\alpha(|x|)}\geq \sigma+\frac{\eta}{\theta\alpha(|x|)},\nonumber
\end{eqnarray}
where $\theta>0$ is the adjusted parameter, and $\eta\geq0$ is a virtual state to be designed. Note that the additive item $\frac{\eta}{\theta\alpha(|x|)}$ increases the upper bound of the comparison threshold of the triggering signal. The dynamic triggering rule can be thus given as below
\begin{eqnarray}
\label{dynamic triggering rule}
t_{i+1}=inf\{t>t_i|\eta + \theta ( \sigma \alpha(| x |) - \gamma(| e |) ) \leq 0\}.
\end{eqnarray}
When the virtual state is designed as $\dot { \eta } = - \zeta(\eta) + \sigma\alpha(| x |) - \gamma(| e |)$, it has been proved in \cite{G14} that the inequality $\eta\geq0$ is satisfied and that both the state $x(t)$ and $\eta$ will converge to the origin asymptotically.

As can be observed from the above reviews, the adjustable range of MIET for static and dynamic triggering mechanism is limited. To improve the implementability of theoretical solution on physical platform, this paper follows the design of a flexible dynamic event-triggered scheme that allows  the variable range to be prescribed to some extend, and ensures a positive MIET independent of intrinsic system states.

\section{Dynamic MIET-Designable Event Triggering Control for Nonlinear Systems}
Through observing the two types of  triggering mechanisms (\ref{static triggering rule}) and (\ref{dynamic triggering rule}) reviewed in the last section, we   find that the derivation of the upper bound of the comparison threshold of the MIET is dependent on the state $x$ and measurement error $e$, and the value of MIET can only be adjusted in a limited range for certain specific plants. In this section, we aim to propose a novel triggering mechanism with designable MIET for general nonlinear systems.
In contrast to the static or dynamic triggering mechanism, the intuitive idea here is to create a triggering signal $Z(t)$, of which the variable range can be freely designed to some extent. We thus adopt the following event triggering rule
\begin{eqnarray}
\label{Triggering rule 1}
t_0&=&0,\nonumber\\
t_{i+1}&=&inf\{t>t_i|Z(t)=0\},
\end{eqnarray}
where $Z(t_i)$ is reset as $\bar Z$ at a triggering instant and $\bar Z>0$ is the design parameter. Here, the variable $Z(t)$ takes a similar role to a countdown variable with an assigned upper bound $\bar Z$. The dynamics $\dot Z(t)=\omega(\varpi(x,e),\varepsilon)$ is considered in the sequel of this paper for designing event triggering conditions, where $\varepsilon>0$ is design parameter.

Thus, the first main result of this paper can be given as follows.
\begin{thm}
\label{theorem1}
Consider the nonlinear  control systems (\ref{discrete closed-loop system}) with the event triggering mechanism (\ref{Triggering rule 1}). The dynamics of additional variable is given as $\dot Z(t)=\omega(\varpi(x,e),\varepsilon)$. Then, for all initial conditions $x(0)$, the closed-loop control system is always guaranteed to converge to the origin asymptotically. Meanwhile, there exists a designable MIET lower bounded by $\tau_1$
\begin{eqnarray}
\label{tao1}
\tau_1 &=&  \sqrt{\frac{1}{b\varepsilon}}\{\mathbf{atan}[\sqrt{\frac{b}{\varepsilon}}(1+\bar Z)]
-\mathbf{atan}[\sqrt{\frac{b}{\varepsilon}}]\}>0, \nonumber\\
b&=&L^{2} \frac{|M|^{2}}{\lambda_{\min }(M)}
\end{eqnarray}
with certain design parameters $M, \varepsilon, \bar Z$ to be detailed in the sequel, for the triggering sequence $(t_i)_{i\rightarrow+\infty}$.
\end{thm}
\begin{pf}
We first analyze the stability. Choose the candidate Lyapunov function as $W =V+\frac{1}{2} Z e^{\top} M e $, where $M$ is a  symmetric positive definite matrix. Note that the derivative of $W$ along the solution of (3) is  $\dot{W} =\dot{V}+\frac{1}{2} \omega e^{\top} M e+Z e^{\top} M \dot{e}$. Because of $\dot{x}=-\dot{e}$ and inequality (\ref{ISS}), it follows
\begin{eqnarray}
\dot{W}&\leq&-\alpha(|x|)+\gamma(|e|)+\frac{1}{2} \omega e^{\top} M e-Ze^{\top} M \dot{x} \nonumber\\
&=&-\alpha(|x|)+\gamma(|e|)+\frac{1}{2} \omega e^{\top} M e-Z e^{\top} M f(x,k(x+e))\nonumber
\end{eqnarray}
Since $\omega<0$, we have
\begin{eqnarray}
\dot{W}&\leq&-\alpha(|x|)+\gamma(|e|)+\frac{1}{2} \omega \lambda_{\min }(M)|e|^{2}\nonumber\\
&&+Z|M||e||f(x,k(x+e))|\nonumber
\end{eqnarray}
Because that the Lipschitz continuity on compact sets of $f(x,u)$ and $k(x)$ renders that $f(x,k(x+e))$ is also Lipschitz continuous, we can thus obtain $|f(x,k(x+e))|\leq L|x|+L|e|$ with Lipschitz constant $L>0$. These facts lead to
\begin{eqnarray}
\dot{W}&\leq&-\alpha(|x|)+\gamma(|e|)+\frac{1}{2} \omega \lambda_{\min }(M)|e|^{2}\nonumber\\
&&+Z|M||e|(L|x|+L|e|)\nonumber\\
&=&-\alpha(|x|)+\gamma(|e|)+\frac{1}{2} \omega \lambda_{\min }(M)|e|^{2}\nonumber\\
&&+ZL|M||e||x|+ZL|M||e|^2\nonumber
\end{eqnarray}
In order to guarantee the asymptotic stability of control system, we enforce the following inequality
\begin{eqnarray}
\omega&<&\frac{2\alpha(|x|)}{\lambda_{\min}(M)} \cdot \frac{1}{|e|^{2}}-\frac{2 \gamma(|e|)}{\lambda_{\min }(M)} \cdot \frac{1}{|e|^{2}}\nonumber\\
&&-\frac{2 L Z|M|}{\lambda_{\min }(M)} \cdot \frac{|x|}{|e|}-\frac{2 L Z|M|}{\lambda_{\min }(M)}.\nonumber
\end{eqnarray}
If more conservative class $K_{+\infty}$ functions $\alpha(|x|)=\frac{1}{2}|x|^{2}$ and $ \gamma(|e|)=L|M||x||e|$ are chosen, the variable $\omega$ further satisfies
\begin{eqnarray}
\omega<\varpi=\frac{1}{\lambda_{\min} (M)} \cdot \frac{|x|^{2}}{|e|^{2}}-(1+Z) \frac{2 L|M|}{\lambda_{\min} (M)} \cdot \frac{|x|}{|e|}\nonumber
\end{eqnarray}
By using the Young inequality, we can obtain
\begin{eqnarray}
-(1+Z) \frac{2 L|M|}{\lambda_{\min }(M)} \cdot \frac{|x|}{|e|} \geq-b(1+Z)^{2}-\frac{1}{b} \cdot \frac{L^{2}|M|^{2}}{\lambda_{\min }^{2}(M)} \cdot \frac{|x|^{2}}{|e|^{2}}\nonumber
\end{eqnarray}
such that
\begin{eqnarray}
\varpi \geq-b(1+Z)^{2}+\left(\frac{1}{\lambda_{\min}(M)}-\frac{1}{b} \cdot \frac{L^{2}|M|^{2}}{\lambda_{\min}^{2}(M)}\right) \cdot \frac{|x|^{2}}{|e|^{2}}\nonumber
\end{eqnarray}
By letting  the second item in the right hand side of the above inequality   satisfy $ b=L^{2} \frac{|M|^{2}}{\lambda_{\min }(M)}$, we can write
\begin{eqnarray}
\varpi-\varepsilon \geq-b(1+Z)^{2}-\varepsilon,\nonumber
\end{eqnarray}
where the design parameter $\varepsilon>0$.
If we further design $\omega$ as
\begin{eqnarray}
\label{dynamics of additional variable}
\omega=\left\{
             \begin{array}{lr}
             min(0,\varpi)-\varepsilon, &e\neq0,  \\
             -\varepsilon,  &e=0,
             \end{array}
\right.
\end{eqnarray}
and consider two cases of $e$: (1)$e\neq0$, if $\varpi<0$, then $\omega=\varpi-\varepsilon$ and if $\varpi\geq0$, $\omega=-\varepsilon\geq-b(1+Z)^{2}-\varepsilon$, so is case (2) when $e=0$. Note that  because $\omega<\varpi$ can be always guaranteed by the design (\ref{dynamics of additional variable}) of $\omega$, we conclude that the Lyapunov function $W$ decreases such that $x(t)$ converges to the origin asymptotically. Moreover,  the dynamics of the countdown variable $Z$ gives the estimate $\dot Z\geq-b(1+Z)^{2}-\varepsilon$. Let $\phi$ be the solution of differential equation $\dot \phi=-b(1+\phi)^{2}-\varepsilon$, then the inter-event time is lower bounded by the time $\tau_1$ that it takes for $\phi$ to evolve from $\bar Z$ to 0. We therefore conclude the formula of $\tau_1$ in (9).
\qed
\end{pf}
We also note that the derivation of the MIET $\tau_1$ does not involve the state $x$ or the measurement error $e$, implying  that the event triggering mechanism developed here has the global robust event-separation property.
\subsubsection{Remark 1.}
There are  only two independent design parameters $\bar Z$ and $\varepsilon$ in the proposed event-triggering scheme. Intuitively, they have the opposite effects on the time interval between two consecutive events. We also note that an event-triggered control system with an strictly positive MIET automatically excludes the Zeno behavior.

\subsubsection{Remark 2.}
Note that there exists an upper bound for the MIET which can be obtained by the following calculation
\begin{eqnarray}
\tau_{1,max}\triangleq\lim_{\varepsilon\to0^{+}}\lim_{\bar Z\to+\infty}\tau_1=\frac{1}{b}\quad when\quad \frac{\varepsilon}{1+\bar Z}>b,\nonumber\\
\tau_{1,max}\triangleq\lim_{\varepsilon\to0^{+}}\lim_{\bar Z\to+\infty}\tau_1=+\infty\quad when\quad \frac{\varepsilon}{1+\bar Z}<b,\nonumber
\end{eqnarray}
Here, we have established an explicit quantitative relation between $\tau_{1,max}$ and $b$, which represents the communication cost and decay rate, respectively.

\section{Dynamic MIET-Designable Event Triggering Control Of General Linear Systems}
In this section, we will specify previous results to event-triggered control of general linear systems with designable MIET.
\subsection{Basic Algorithm}
Consider a general linear  control system of the form
\begin{eqnarray}
\label{general linear systems}
\dot x(t)=Ax(t)+Bu(t),
\end{eqnarray}
where $A$ and $B$ are system matrices with proper dimensions, and we assume the system is controllable.
A feedback control law is designed as $u(t)=Kx(t)$ through pole assignment for stabilizing the system (\ref{general linear systems}). The  closed-loop control system is then obtained as below
\begin{eqnarray}
\dot x(t)=(A+BK)x(t).\nonumber
\end{eqnarray}
This thus implies that there exists a Lyapunov function $V=x^{\top}Px$ such that the symmetric positive definite matrix $P$ satisfies
\begin{eqnarray}
(A+B K)^{\top} P+P(A+B K)=-Q,\nonumber
\end{eqnarray}
where $Q$ is an arbitrary symmetric positive definite matrix.
When the state-feedback control law $u(t)=Kx(t)$, which is updated in a continuous time manner, is conducted on digital platforms and/or wireless communication environment, then it needs to be modified as discrete-time updates. In this section, we formally propose an MIET-designable event triggering method to schedule the computation and communication resources and determine the triggering time that updates the feedback of the system states $x(t)$ into the closed-loop control system; i.e., the control is modified as $u(t)=Kx(t_i), t\in[t_i, t_{i+1})$, where $t_i$ is the triggering instant.

Following the idea of the event-triggered control framework presented in Section 2, we define the measurement error $e(t)=x(t_i)-x(t), t\in[t_i, t_{i+1})$; the following closed-loop system is thus rendered
\begin{eqnarray}
\label{closed loop event triggering linear system}
\dot x(t)=Ax(t)+BKx(t)+BKe(t).
    \end{eqnarray}
Similar to the event triggering mechanism (\ref{Triggering rule 1}), we apply the same triggering rule for the general linear system, and define the additional event function dynamics as $\dot Z(t)=\omega(\varpi,\varepsilon)$.

We can now give the second main contribution of paper.
\begin{thm}
\label{theorem2}
Consider the general linear  control system  (\ref{closed loop event triggering linear system}) with the event triggering mechanism (\ref{Triggering rule 1}) for all initial condition $x(0)$. Then, $x(t)$ asymptotically converge to the origin. Meanwhile, there exists a designable MIET lower bounded by $\tau_2$
\begin{eqnarray}
\label{tao2}
\tau_2 &=&  \sqrt{\frac{1}{b\varepsilon}}\{\mathbf{atan}[\sqrt{\frac{b}{\varepsilon}}(1+\bar Z)]
-\mathbf{atan}[\sqrt{\frac{b}{\varepsilon}}]\}>0, \nonumber\\
b &=& \frac{|P B K|^{2}}{\lambda_{\min }(P) \lambda_{\min }(Q)}
\end{eqnarray}
for the triggering sequence $(t_i)_{i\rightarrow+\infty}$.
\end{thm}
\begin{pf}
We choose a  Lyapunov function candidate $W=\frac{1}{2} x^{\top} P x+\frac{1}{2}Ze^{\top}Pe$. Its derivative along the solution of (12) gives
\begin{eqnarray}
\dot{W}&=&x^{\top} P \dot{x}+\frac{1}{2} \omega e^{\top} P e+Ze^{\top} P \dot{e}\nonumber\\
&=&-\frac{1}{2} x^{\top} Q x+x^{\top} P B K e\nonumber\\
&&+\frac{1}{2} \omega e^{\top} P e-Z e^{\top} P(A+B K) x\nonumber\\
&&-Z e^{T} P B K e\nonumber\\
&\leq&-\frac{1}{2} \lambda_{\min}(Q)|x|^{2}+|P B K||x||e|\nonumber\\
&&+\frac{1}{2} \omega \lambda_{\min}(P)|e|^{2}+Z|PA||x||e|\nonumber\\
&&+Z|PBK||x||e|+Z|PBK||e|^2\nonumber
\end{eqnarray}
By enforcing the following inequality
\begin{eqnarray}
\omega&<&\varpi=\frac{\lambda_{\min }(Q)}{\lambda_{\min }(P)} \frac{|x|^{2}}{|e|^{2}}-2 \frac{|P BK|}{\lambda_{\min }(P)} \frac{|x|}{|e|}-2 Z \frac{|P B K|}{\lambda_{\min} (P)} \frac{|x|}{|e|},\nonumber\\
&=& \frac{\lambda_{\min }(Q)}{\lambda_{\min }(P)} \frac{|x|^{2}}{|e|^{2}}-2(1+Z) \frac{|P B K|}{\lambda_{\min }(P)} \frac{|x|}{|e|},\nonumber
\end{eqnarray}
we can guarantee the asymptotic stability of the closed-loop system (\ref{closed loop event triggering linear system}).
Analogous to the nonlinear case, through some manipulations based on the Young inequality, it follows that
\begin{eqnarray}
\varpi &\geq& -b(1+Z)^{2}+\left(\frac{\lambda \min (Q)}{\lambda_{\min} (P)}
-\frac{|P B K|^{2}}{b\lambda_{\min}^{2}(P)}\right) \frac{|x|^{2}}{|e|^{2}}.\nonumber
\end{eqnarray}
When we choose $b$ to satisfy $ b= \frac{|P B K|^{2}}{\lambda_{\min }(P) \lambda_{\min }(Q)}$, then $\varpi-\varepsilon \geq-b(1+Z)^{2}-\varepsilon$ can be obtained where $\varepsilon>0$ is a design parameter. Recalling the definition of $\omega$ in (\ref{dynamics of additional variable}) for all value of $\varpi$, we can always write $\omega \geq-b(1+Z)^{2}-\varepsilon$ if $e\not=0$. Note that the inequality is also satisfied if $e=0$ and $\omega=-\varepsilon$. Therefore, we can conclude that $\phi\leq Z$ where $\phi$ is the solution of $\dot \phi \geq-b(1+\phi)^{2}-\varepsilon$ satisfying $\phi(0)=\bar Z$. The MIET (\ref{tao2}) can be thus lower bounded by the time $\tau_2$ that it takes to reach $\phi(\tau_2)=0$, which gives the explicit solution as in (13). Moreover, for any initial state $x(0)$ the Lyapunov function $W$ converges to the origin asymptotically due to the fact $\omega<\varpi$. \qed
\end{pf}

\subsubsection{Remark 3.}
In Theorem \ref{theorem2}, a freely designable MIET only relying on design parameters $\bar Z, \varepsilon$ is completely established. Moreover, we can further enlarge the positive MIET by decreasing the parameter $b$, which can be realized through adjusting the matrix $Q$ and $P$, and the control gain $K$. Analogous to the case of nonlinear system, the event triggering mechanism also holds the global robust event-separation property for general linear systems.
\subsubsection{Remark 4.}
Theorems 1 and 2 both give the estimate $\dot \phi =-b(1+\phi)^{2}-\varepsilon$ of the same form, and $\tau_1=\tau_2$. It is worth noting that this estimate also matches the results in \cite{NGC19}, which implies the similar event triggering mechanism with closely related properties.

\subsection{Robustness Analysis}
We continue to consider the following perturbed linear system
\begin{eqnarray}
\label{add disturbance}
\dot x(t)=Ax(t)+Bu(t)+Hd(t).
\end{eqnarray}
\begin{proposition}
\label{theorem3}
Consider the general linear  control systems (\ref{add disturbance}) with the event triggering mechanism (\ref{Triggering rule 1}) and any bounded disturbance $|d|\leq\bar d$ for all initial condition $x(0)$. Then there exists the same positive lower bound for the designable MIET $\tau_3=\tau_2$, implying that the minimum event interval is robust to perturbations. Furthermore, suppose that the perturbation $d(t)$ is convergent. Then the system state also asymptotically converges to the origin.
\end{proposition}
\begin{pf}
We first analyze the robustness issues of the algorithm. Recall the Lyapunov function candidate and its derivation
\begin{eqnarray}
\dot W&\leq&-\frac{1}{2} \lambda_{\min}(Q)|x|^{2}+|P B K||x||e|+\frac{1}{2} \omega \lambda_{\min}(P)|e|^{2}\nonumber\\
&&+Z|PA||x||e|+Z|PBK||x||e|+Z|PBK||e|^2\nonumber\\
&&+|PH||x|\bar d+Z|PH||e|\bar d\nonumber
\end{eqnarray}
Therefore, it is clear that the formula below is still satisfied
\begin{eqnarray}
\omega<\varpi= \frac{\lambda_{\min }(Q)}{\lambda_{\min }(P)} \frac{|x|^{2}}{|e|^{2}}-2(1+Z) \frac{|P B K|}{\lambda_{\min }(P)} \frac{|x|}{|e|}.\nonumber
\end{eqnarray}
Following the same lines in the proof of Theorem 2,  it can be found that its derivation process only depends on the design parameters $\bar Z, b, \varepsilon$. We therefore conclude that the minimum event-interval is still guaranteed in the presence of system perturbation $d(t)$. The convergent $d(t)$ implying convergent $x(t)$ is a consequence of the exponential stability of the linear system.  \qed

\end{pf}

\section{Numerical simulations}
In this section, two simulations are given to show the effectiveness of the proposed theoretical results.
\subsection{Nonlinear system}
Firstly, let us consider the forced van der Pol oscillator: $\dot x_1=x_2$, $\dot x_2=(1-x^2_1)x_2-x_1+u$, where $x_1,x_2\in\mathbb{R}$ are states and $u$ is the control input to be designed. Here, we adopt the control law $u=-x_2-(1-x^2_1)x_2$ which can stabilize the origin of the system. The nominal values of design parameters are chosen as $\bar Z=1$, $\varepsilon=1$. The Lipschitz constant $L$ and matrix $M$ are obtained as $1$ and $[1\quad0.25;0.25 \quad1]$, respectively. The initial states are arbitrarily set as $x_1=1, x_2=-0.5$.

\begin{figure}[h]
\label{shiyitu}
    \centering
    \subfigure[]{\includegraphics[width=3.2in,height=2in]{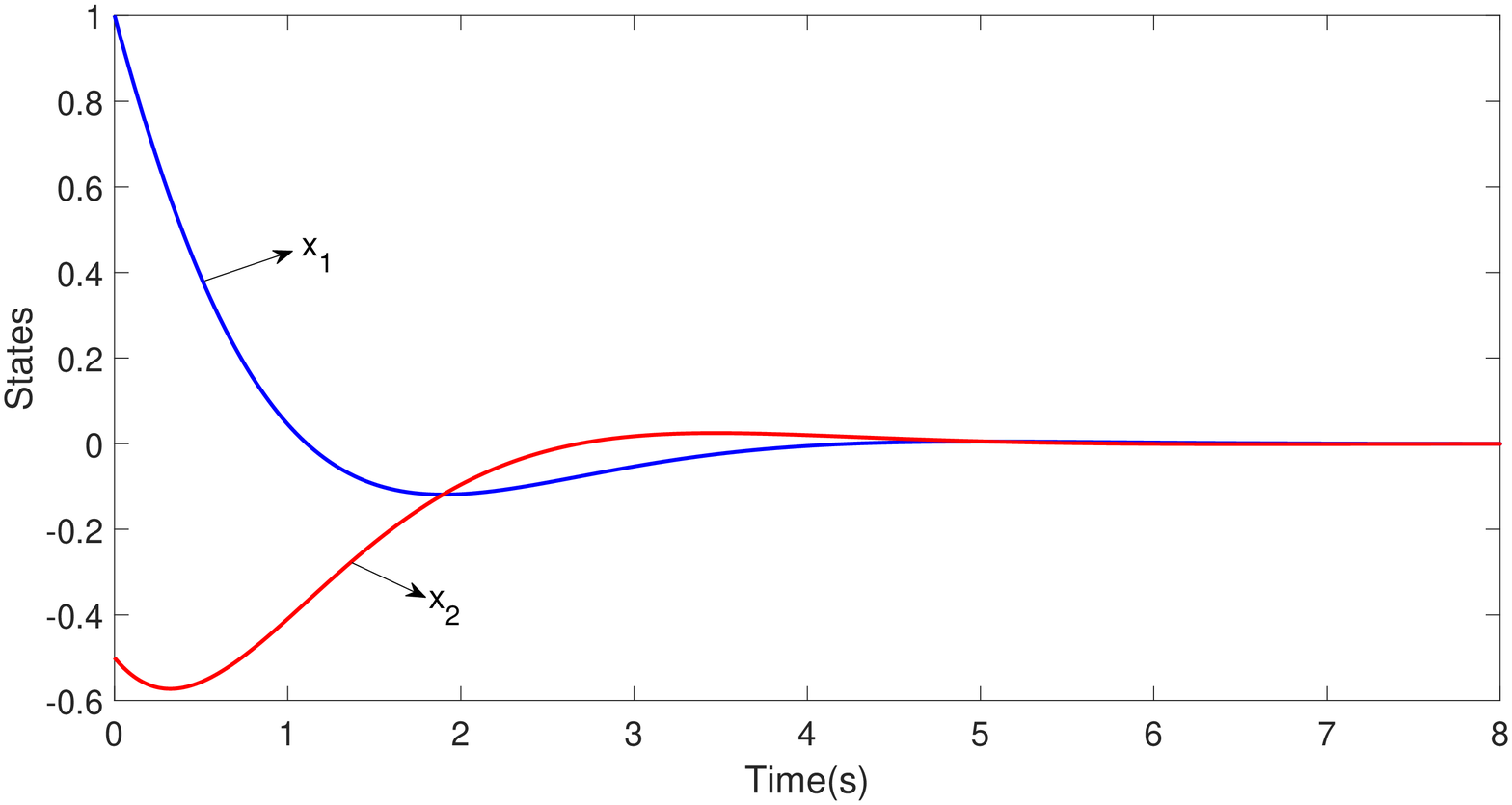}}
    \subfigure[]{\includegraphics[width=3.2in,height=2in]{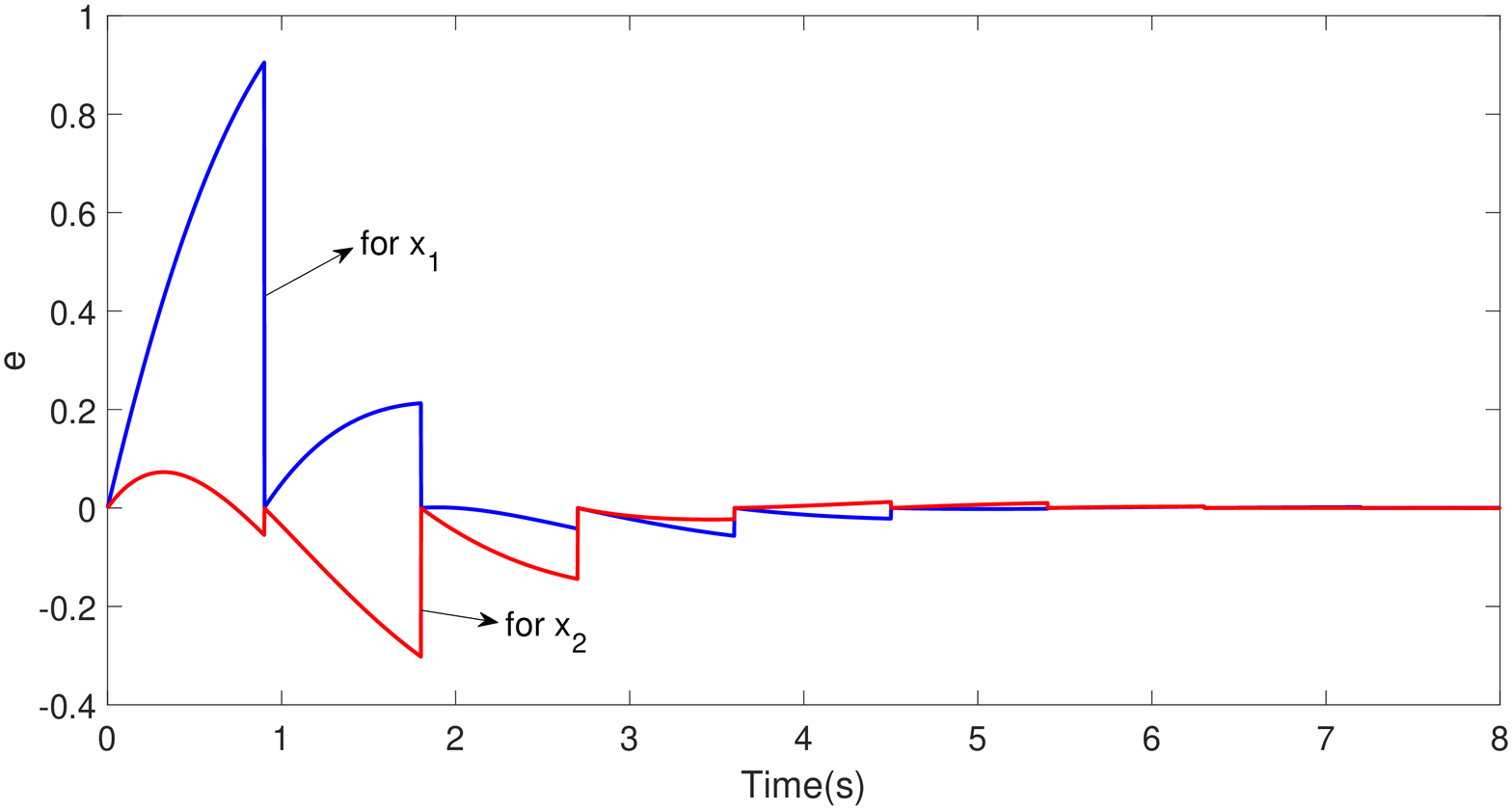}}
    \caption{(a) The trajectories of states $x_1, x_2$, (b) the trajectories of the measurement errors $e_1, e_2$, both for event-triggered nonlinear system}
\end{figure}

\begin{figure}[h]
\label{shiyitu}
    \centering
    \includegraphics[width=3.2in,height=2in]{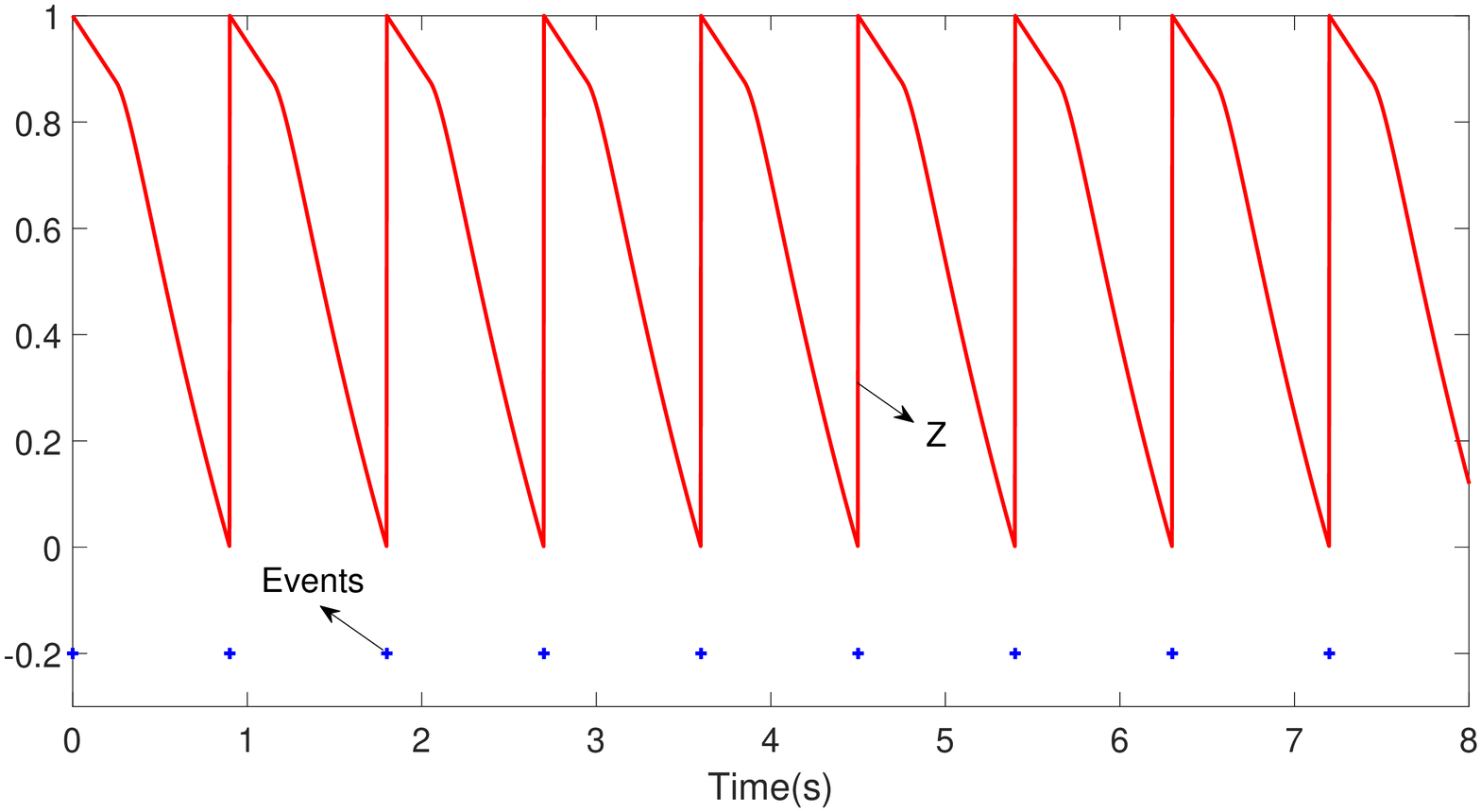}
    \caption{The evolution of the additional dynamic variable $Z$ and the related events for event-triggered nonlinear system}
\end{figure}
In Fig.1, the asymptotic convergence of states $x_1, x_2$ and measurement error $e$ validates the related results in Theorem 1. The evolution of the additional dynamic variable $Z$ can be observed in Fig.2. It shows that the events are triggered almost periodically for every 0.9$s$. Note that this is different to the sampled-data control with a fixed period where the system is open-loop control between two continuous sampling instants. In the proposed event-triggering framework, we can increase the design parameter $\bar Z$ so as to reach a larger inter-event time while guaranteeing the stability. For example, the interval between two consecutive events increases up to 3.722$s$ when the parameter $\bar Z=3$ is chosen. Also, after we have the matrix $M$ and Lipschitz constant $L$, $b=2.083$ can be directly obtained because of equation (\ref{tao1}). In this case, the positive lower bound $\tau_1$ can be calculated as 0.189$s$, which is less than the practical simulation result.
\subsection{Linear system}
In order to compare the present results with the static and dynamic triggering mechanisms for linear systems in \cite{T07} and \cite{G14}, we use the same linear plant model with the same controller and choose the same gains. Specifically, choosing $A=[0\quad1;-2 \quad3]$, $B=[0;1]$, $K=[1 \quad -4]$ can get $P=[1\quad 1/4;1/4\quad 1]$ and $Q=[1/2 \quad1/4;1/4 \quad3/2]$. Since $\frac{|P B K|^{2}}{\lambda_{\min }(P) \lambda_{\min }(Q)}=54.61$, based on formula (\ref{tao2}) we choose $b=55$. Meanwhile, by adopting  the same simulation setup in \cite{JC19}, $\bar Z=1$ and $\varepsilon=1$ are chosen as the design parameters. We set the initial states as $x_1=10, x_2=0$.

\begin{figure}[h]
\label{shiyitu}
    \centering
    \subfigure[]{\includegraphics[width=3.2in,height=2in]{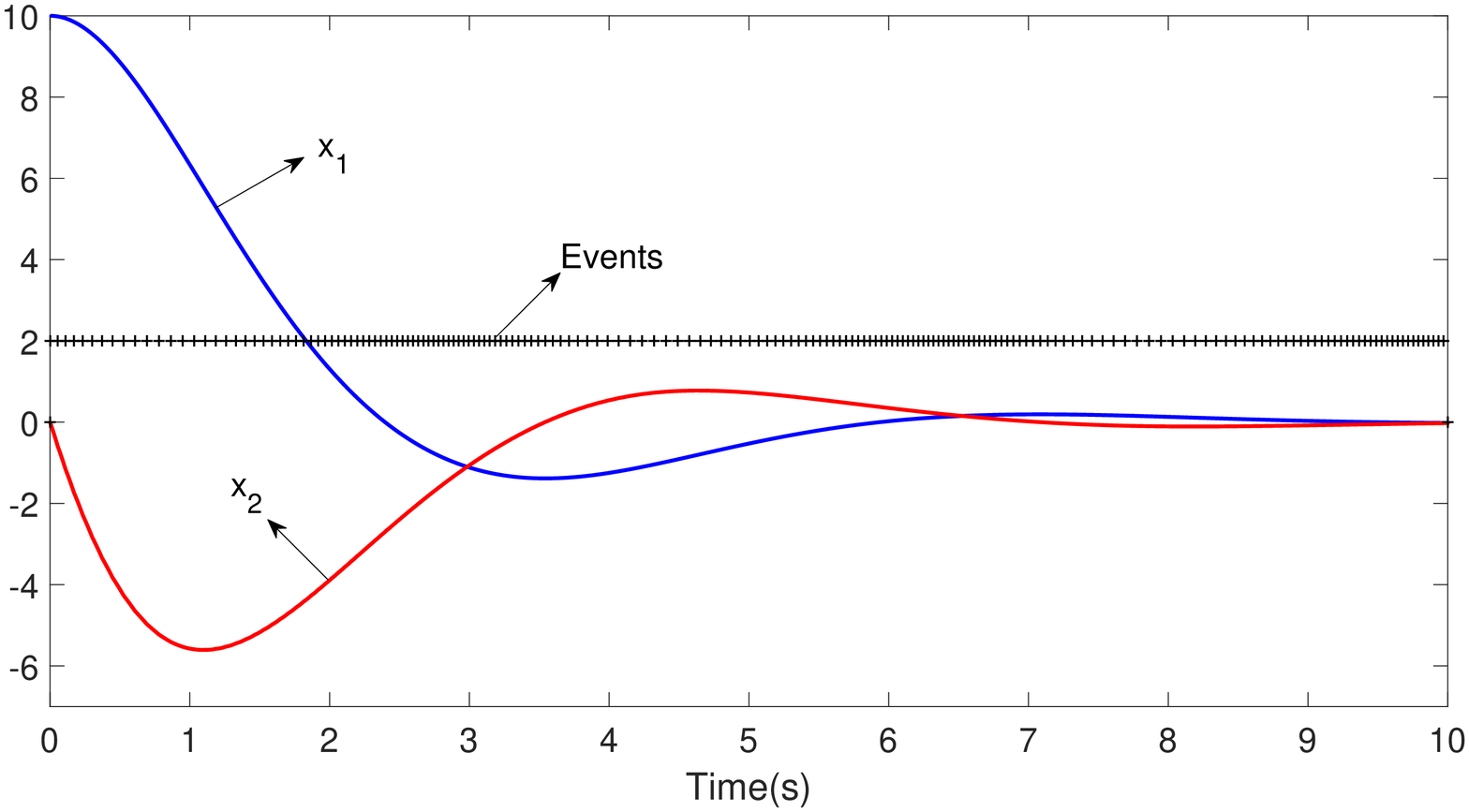}}
    \subfigure[]{\includegraphics[width=3.2in,height=2in]{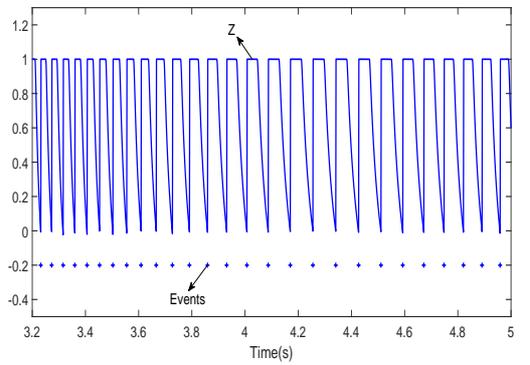}}
    \caption{Numerical simulation results of the dynamic MIET-designable event-triggered linear control system with $\bar Z=1$ and $\varepsilon=1$. (a) The trajectories and event triggering time, respectively. (b) The triggering events and the evolution of $Z$ from 3.2 to 5s.}
\end{figure}

\begin{figure}[h]
\label{shiyitu}
    \centering
    \includegraphics[width=3.2in,height=2in]{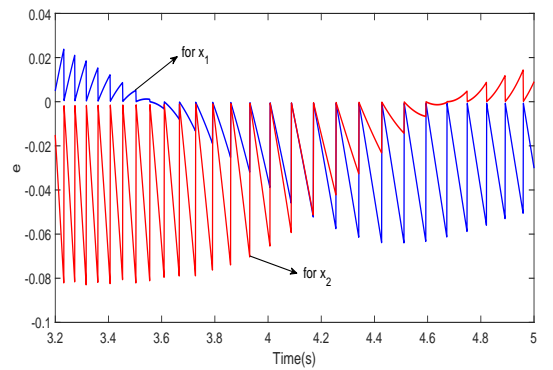}
    \caption{The evolution of the measurement errors $e_1, e_2$ for event-triggered linear control system.}
\end{figure}

From Fig.3 and Fig.6, it can be found that the states $x$ and measurement error $e$ converge to the origin asymptotically which validates the stability of the general linear control system with the MIET-designable event triggering mechanism. Based on the formula (\ref{tao2}), we can calculate a lower bound of MIET as 9 $ms$, which is smaller than the simulation result of 36 $ms$, implying that the calculated lower bound MIET may be conservative.  In the meantime, from the simulation, we can compute the maximum inter-event time as 86 $ms$.
\begin{figure}[h]
\label{shiyitu}
    \centering
    \includegraphics[width=3.2in,height=2in]{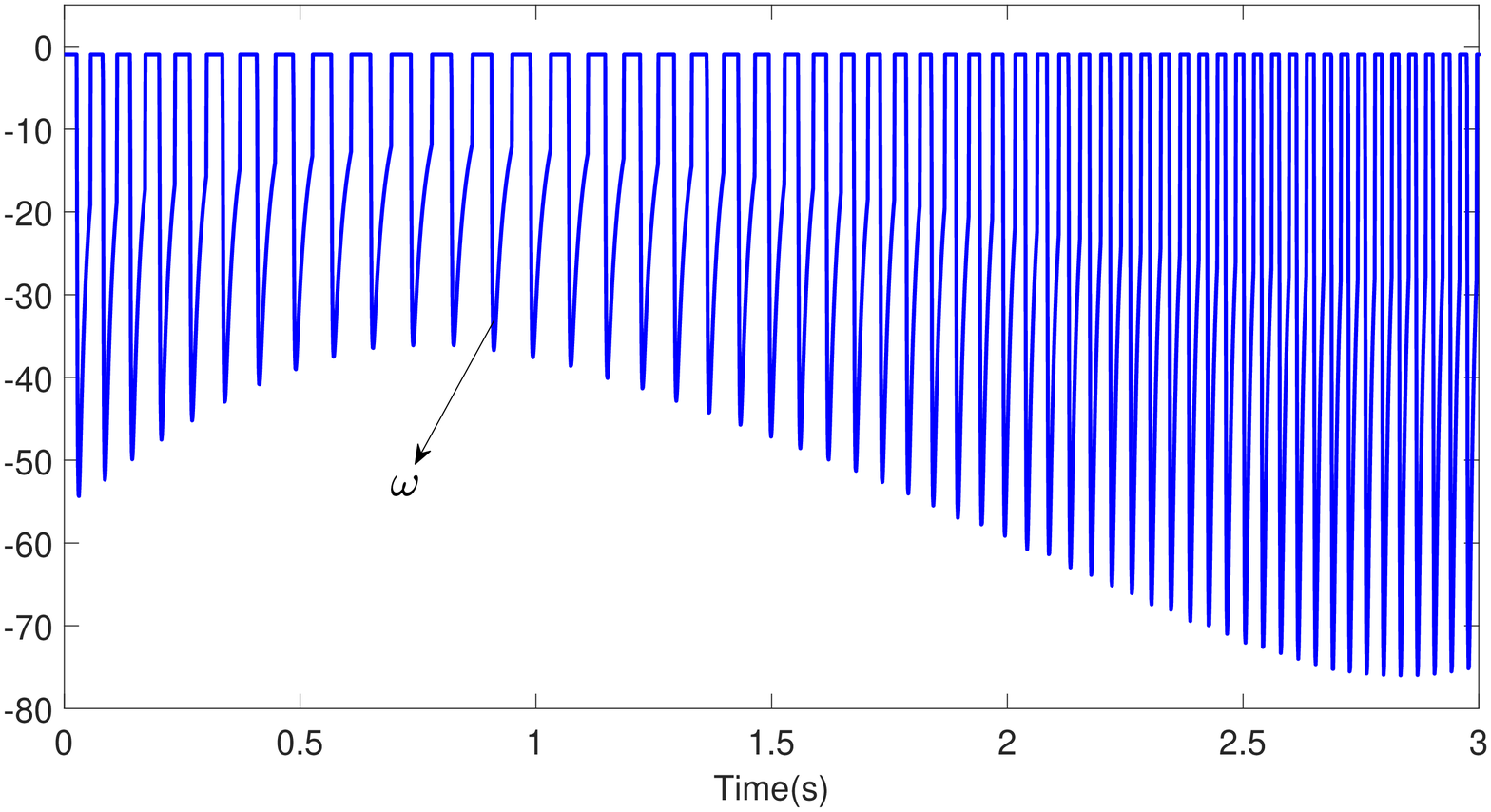}
    \caption{The evolution of $\omega$ for event-triggered linear control system.}
\end{figure}
The variable $\omega$ always keeps smaller than $-\varepsilon$ in Fig.5, which implies that the clock-like variable $Z$ always decreases while the speed rate is changed throughout the whole countdown process. The evolution of the term ``$\frac{1}{2}Ze^{\top}Pe$" is shown in Fig.6, which is not monotonous.
\begin{figure}[h]
\label{shiyitu}
    \centering
    \includegraphics[width=3.2in,height=2in]{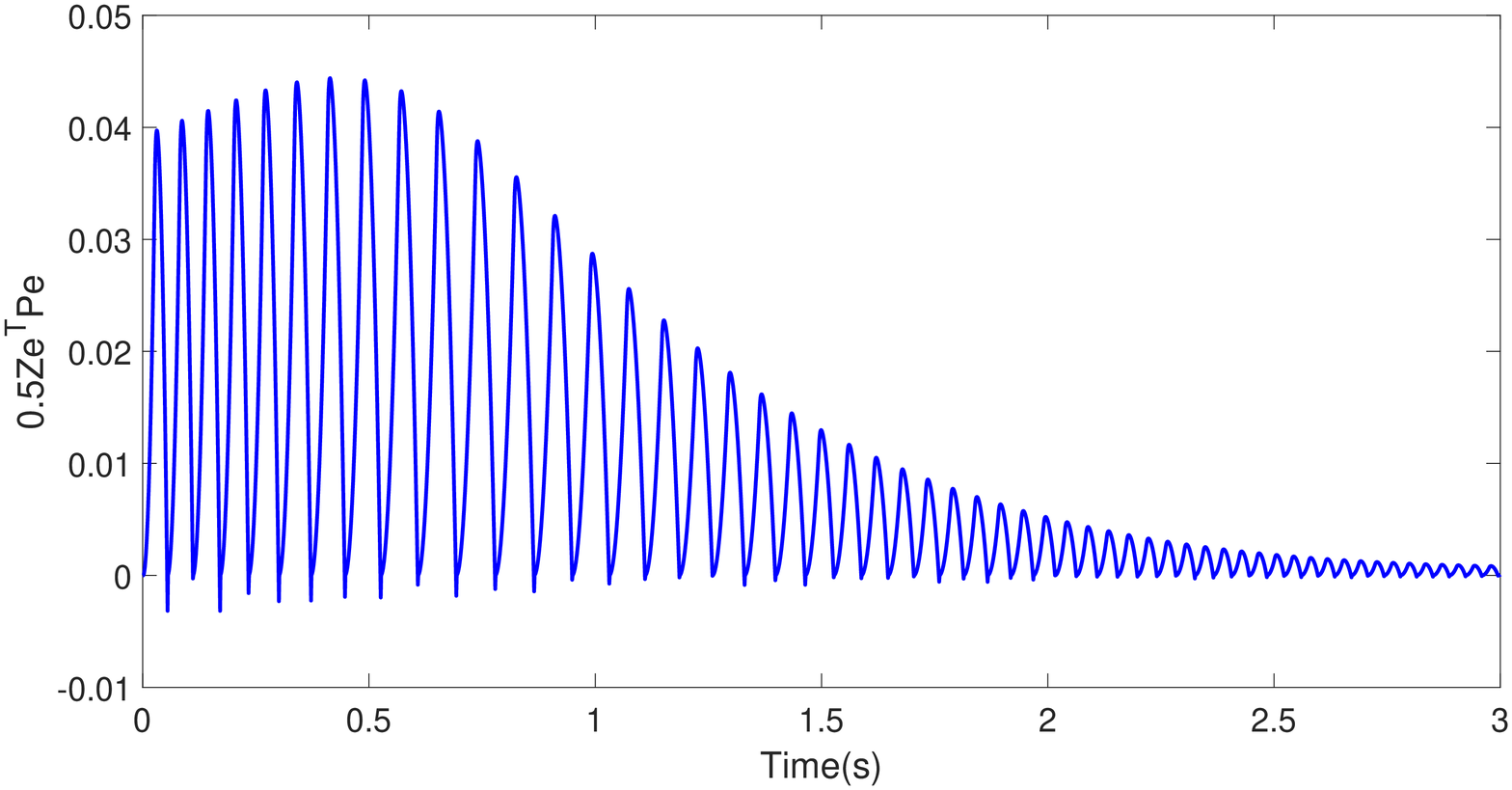}
    \caption{The evolution of the item $\frac{1}{2}Ze^{\top}Pe$ in the Lyapunov function.}
\end{figure}
Next, we compute the eigenvalues of state matrix $A+BK$ in \cite{PSH19} as $\lambda_1=-0.5+0.866j$ and $\lambda_2=-0.5-0.866j$, which are complex conjugates and non-real. Furthermore, it is noticed that $\pi/0.866=3.6277$ is very close to the period observed in Fig.7. All of these facts are consistent with the Theorem 3 in \cite{PSH19}, in which the planar linear system and static event condition $\|\hat{x}(t)-x(t)\|\leq\sigma\|x(t)\|$ in \cite{T07} are considered. Moreover, initial states $[10;0],[-10;0],[0;10],[0;-10],[5;5]$ are implemented in the same setting, and the results further validate the statement in Theorem 3. It is shown that the period of inter-event times is irrelevant to the initial state of controlled system, but might lead to different phase. In the meantime, these findings also provide some hints for the connection between static and dynamic triggering mechanisms.
\begin{figure}[h]
\label{shiyitu}
    \centering
    \includegraphics[width=3.2in,height=2in]{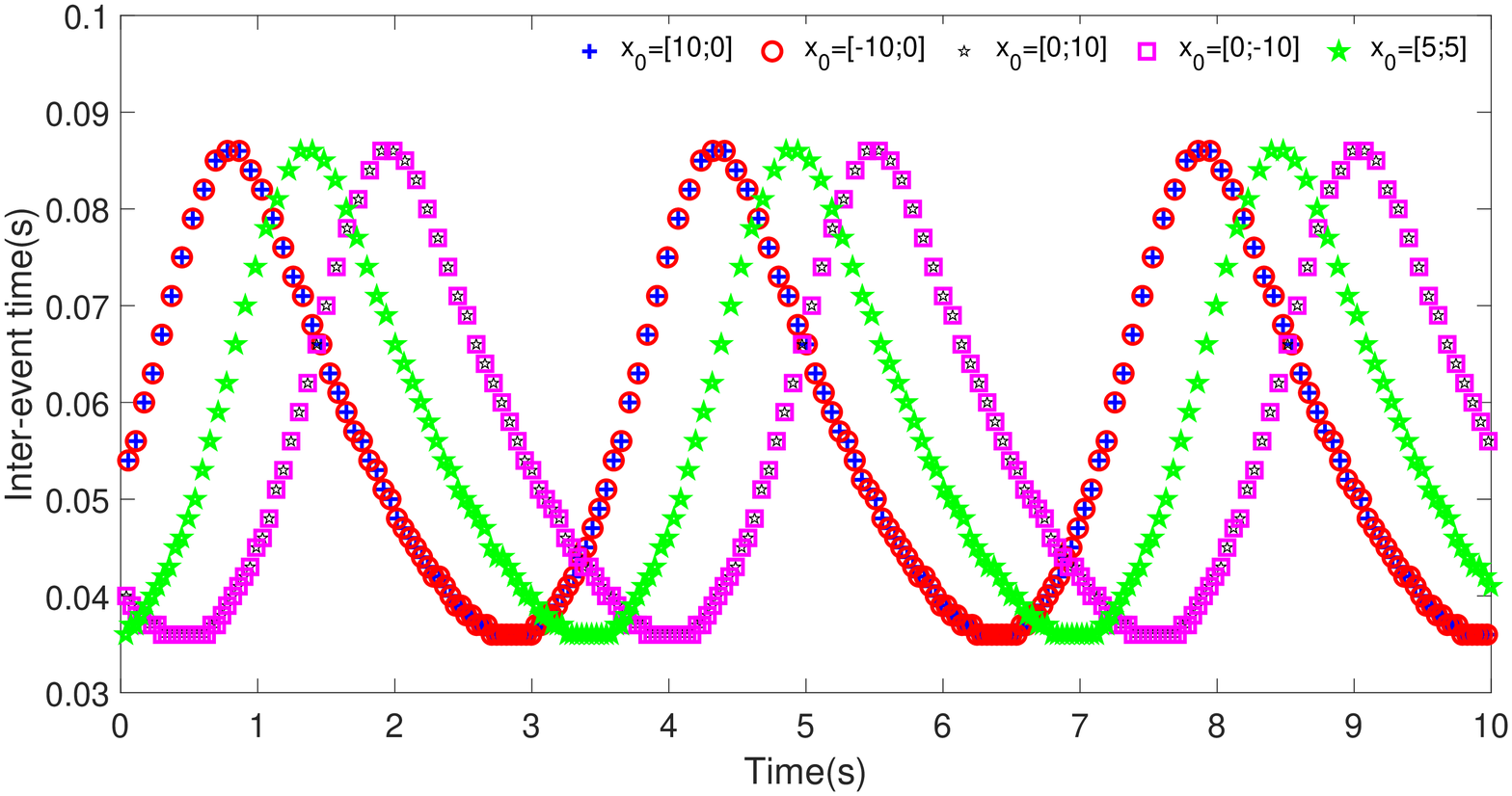}
    \caption{The evolution of the inter-event times.}
\end{figure}
\section{Concluding remarks}
In this work, in order to improve certain crucial characteristics like MIET and event-separation property, we develop a new framework of the design and analysis of the event-triggered control system. A MIET-designable triggering mechanism has been established for nonlinear system and general linear system, respectively. Afterwards, the robustness issue of the proposed results is further considered. It is shown that the present MIET-designable triggering mechanism guarantees Zeno-free triggering and the robust global event-separation property.

Currently, we are working on applying the proposed triggering mechanism to the distributed control and networked control systems with general linear systems. It is also interesting to investigate other kinds of disturbances, such as the time delay, DoS attack, or timing error, etc. In the future, we also plan to apply the MEIT-designable event triggering mechanism to more broad fields, such as cyber-physical systems and power network systems.

\bibliographystyle{ieeetr}
\bibliography{ifacconf}

\end{document}